\documentclass[final,3p,times,twocolumn,sort&compress]{elsarticle}
\usepackage{amssymb}
\usepackage{hyperref}
\usepackage[tight, nice]{units}
\usepackage{color}
\usepackage{amsmath}
\journal{NIM A}
\begin{document}
\begin{frontmatter}
\title{Vertical Beam Polarization at MAMI}
\author[kph]{B.\,S.~Schlimme\corref{CorrespondingAuthor}}
\ead{schlimme@kph.uni-mainz.de}
\ead[url]{www.kph.uni-mainz.de}
\cortext[CorrespondingAuthor]{Corresponding author.}
\author[kph]{P.~Achenbach}
\author[kph]{K.~Aulenbacher}
\author[kph]{S.~Baunack}
\author[kph]{D.~Bender}
\author[stefan]{J.~Beri\v{c}i\v{c}}
\author[zagreb]{D.~Bosnar}
\author[clermont]{L.~Correa}
\author[kph]{M.~Dehn}
\author[kph]{M.\,O.~Distler}
\author[kph]{A.~Esser}
\author[clermont]{H.~Fonvieille}
\author[zagreb]{I.~Fri\v{s}\v{c}i\'c\fnref{IF}}
\fntext[IF]{Present address: MIT-LNS, Cambridge MA, 02139, USA}
\author[kph]{B.~Gutheil}
\author[kph]{P.~Herrmann}
\author[kph]{M.~Hoek}
\author[kph]{S.~Kegel}
\author[kph]{Y.~Kohl}
\author[stefan]{T.~Kolar}
\author[kph]{H.-J.~Kreidel}
\author[kph]{F.~Maas}
\author[kph]{H.~Merkel}
\author[kph]{M.~Mihovilovi\v{c}}
\author[kph]{J.~M\"uller}
\author[kph]{U.~M\"uller}
\author[kph]{F.~Nillius}
\author[kph]{A.~Nuck}
\author[kph]{J.~Pochodzalla}
\author[kph]{M.~Schoth}
\author[kph]{F.~Schulz}
\author[kph]{C.~Sfienti}
\author[ljubljana,stefan]{S.~\v{S}irca}
\author[kph]{B.~Spruck}
\author[stefan]{S.~\v{S}tajner}
\author[kph]{M.~Thiel}
\author[kph]{V.~Tioukine}
\author[kph]{A.~Tyukin}
\author[kph]{A.~Weber}
\address[kph]{Institut f\"{u}r Kernphysik, Johannes Gutenberg-Universit\"{a}t Mainz, D-55099 Mainz, Germany}
\address[stefan]{Jo\v{z}ef Stefan Institute, SI-1000 Ljubljana, Slovenia}
\address[zagreb]{Department of Physics, University of Zagreb, HR-10002 Zagreb, Croatia}
\address[clermont]{Clermont Universit\'e, Universit\'e Blaise Pascal, F-63000 Clermont-Ferrand, France}
\address[ljubljana]{Department of Physics, University of Ljubljana, SI-1000 Ljubljana, Slovenia}
\begin{abstract}
For the first time a vertically polarized electron beam has been used for physics experiments at MAMI in the energy range between 180 and 855\,MeV. The beam-normal single-spin asymmetry $A_{\mathrm{n}}$, which is a direct probe of higher-order photon exchange beyond the first Born approximation, has been measured in the reaction $^{12}\mathrm C(\vec e,e')^{12}\mathrm C$. Vertical polarization orientation was necessary to measure this asymmetry with the existing experimental setup. In this paper we describe the procedure to orient the electron polarization vector vertically, and the concept of determining both its magnitude and orientation with the available setup. A sophisticated method has been developed to overcome the lack of a polarimeter setup sensitive to the vertical polarization component.
\end{abstract}
\end{frontmatter}
\section{Introduction}\label{sec:introduction}
Electron scattering is an excellent tool to study the structure of nucleons and atomic nuclei. In addition to conventional observables like unpolarized scattering cross sections, polarization observables allow novel access to relevant quantities. In particular, the use of longitudinally polarized electron beams turns out to be very beneficial, and dedicated experiments have been performed at various places. At the Mainz Microtron (MAMI) electron accelerator \cite{Herminghaus:1976mt,Kaiser:2008zza,Dehn:2011za}, for instance, the electric to magnetic form factor ratios were measured for the proton as well as for the neutron (see \cite{Pospischil:2001pp,Schlimme:2013eoz} and references therein). Moreover, parity-violating (PV) single-spin asymmetries in elastic electron scattering were measured to investigate the distribution of strange quarks within the nucleon \cite{Baunack:2009gy}. In these PV experiments, possible small transverse beam polarization components can lead to significant background through contributions from the {\it beam-normal} single-spin asymmetry $A_\mathrm{n}$. Therefore, the measurement of $A_{\mathrm n}$ is mandatory in order to constrain the systematic error. $A_{\mathrm n}$ itself provides an interesting challenge for theoretical predictions, because this observable is a direct probe of multi-photon exchange processes.

A recent measurement of $A_{\mathrm n}$ in elastic scattering off $^{208}\mathrm{Pb}$ revealed a significant disagreement between the data and the prediction \cite{Abrahamyan:2012cg}. The sources of the disagreement are presently not understood and motivate more measurements for target nuclei in the intermediate mass range. In this context, $A_{\mathrm n}$ of carbon has been recently measured at MAMI by using the spectrometer setup of the A1 collaboration \cite{An2016}.

To observe $A_{\mathrm n}$, the beam polarization vector must have a component normal to the scattering plane, which is spanned by the incident and scattered electron momenta. Because the A1 spectrometers accept horizontally scattered electrons, the beam polarization must be set vertical for maximum sensitivity.

Although MAMI does not provide vertical beam polarization in conventional operation, it was possible to use the existing setup -- with minor modifications -- to run, for the first time, a high energy vertically polarized beam over several weeks. The direction of the polarization was controlled by a combination of two beam line instruments, a Wien filter spin rotator and a double solenoid. However, an accurate determination of the vertical polarization component was challenging, since a polarimeter sensitive to this polarization component was not available. Hence, we investigated a method to deduce the vertical polarization component from a determination of the total beam polarization and the residual horizontal polarization components.

The total beam polarization was measured by using a M\o ller polarimeter with the beam polarization oriented longitudinally in the experimental hall. After rotation of the polarization vector in the vertical direction, the same polarimeter was used to measure the residual longitudinal polarization component, which should vanish for perfect vertical polarization alignment. The transverse horizontal component could be evaluated by measuring again with the M\o ller polarimeter, but at a different beam energy.

Besides dedicated measurements to estimate the contribution from false asymmetries to the M\o ller polarimeter results, we have also performed polarization measurements with Mott and Compton polarimeters, which are located close to the electron source. A careful comparison provides a powerful test of our understanding of systematic effects.

The paper is organized as follows. We briefly review the operating principle of MAMI and its polarized electron source in Sec. \ref{sec:PKA}. 
In Sections \ref{sec:PolRotationHorizontal} and \ref{sec:PolRotationVertical} the precession of the polarization vector in the bending fields of the accelerator as well as in the fields of the Wien filter and the double solenoid is described. The results from the alignment procedure at the beginning of the experiment are shown in Sec. \ref{sec:PolMeasConcept}. Supplementary polarimeter data dedicated to systematic error estimates of the individual polarimeter measurements are evaluated in Sec. \ref{sec:Systematics}. Finally, the data from different measurements are combined in Sec. \ref{sec:OfflineResidualPolarization} and the long-term stability of the polarization alignment is considered in Sec. \ref{sec:HorizontalResultAndStability}. A summary follows in Sec. \ref{sec:Summary}.
\section{The polarized electron source at MAMI}\label{sec:PKA}
The normal-conducting continuous-wave electron accelerator MAMI consists of a cascade of three racetrack microtrons followed by a harmonic double-sided microtron. The source of polarized electrons \cite{Aulenbacher:1997zy, Aulenbacher:2011zz} provides longitudinally polarized electrons. They are produced by using a strained GaAs/GaAsP superlattice photocathode illuminated with circularly polarized laser light. The helicity of the electron beam is selected by setting the high voltage polarity of a Pockels cell. Dedicated tests have been performed to unambiguously determine the absolute sign of the helicity at the source by examining the asymmetries of Mott and M\o ller polarimeter measurements. By insertion of a $\lambda/2$ plate between the laser system and the photocathode, the polarization orientation of the electron beam can be reversed. This allows a test of the understanding of systematic effects \cite{Maas:2004ta}. The measurements presented in the following have been taken without the $\lambda/2$ plate being inserted unless specified otherwise.

\begin{figure}
  \centering
  \includegraphics[width=\linewidth]{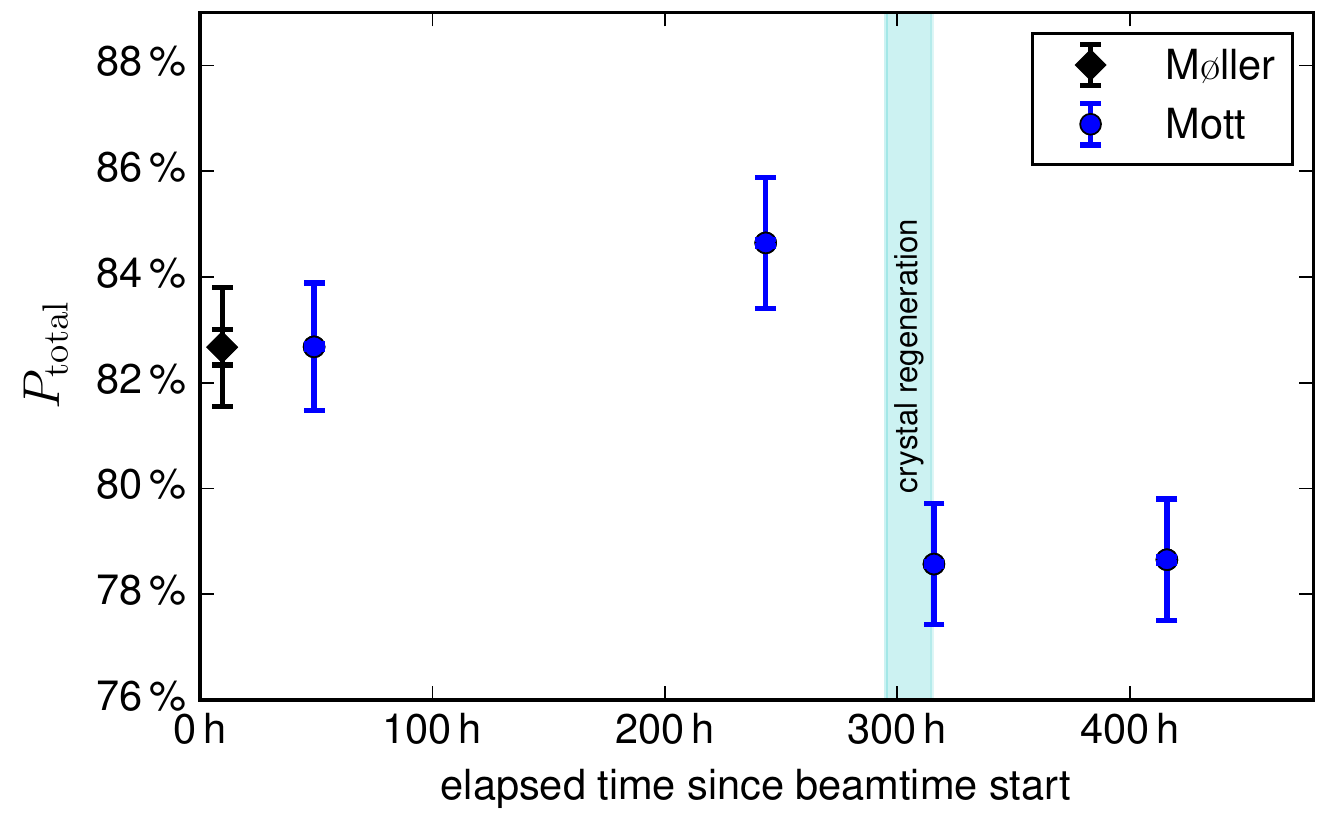}
  \caption{Measurement of the total beam polarization. The absolute value was measured with the M\o ller polarimeter. The Mott polarimeter was used to monitor the usual long-term increase of the polarization, and the drop of the polarization after a necessary regenerating procedure applied to the source crystal.} 
  \label{fig:PTotal}
\end{figure}

The total beam polarization was measured with a M\o ller polarimeter (see \cite{BartschPhD,Schlimme:2012xga,Tyukin:2015master} and references therein) at the beginning of the beamtime. The measurements were performed at a beam energy of $\unit[855]{MeV}$, because systematic errors related to detector acceptances and signal amplitudes are minimal for this energy. The electron polarization vector was aligned longitudinally (compare Sec. \ref{sec:PolRotationHorizontal}) in the spectrometer hall to maximize the sensitivity. The result is $P=82.7\%\pm 0.3\%(\mathrm{stat})\pm 1.1\%(\mathrm{syst})$. To monitor the beam polarization throughout the beamtime, we performed several measurements with a Mott polarimeter \cite{Tioukine:2011Mott,Tioukine:2013hna} for transverse horizontal alignment of the polarization. The absolute scale of the Mott polarimeter measurements was not calibrated before the experiment, therefore we rescaled the results by a factor $1.018$ to match the first Mott polarimeter result with the M\o ller polarimeter result. The results are shown in Fig. \ref{fig:PTotal}.
\section{Rotation of the polarization vector in the horizontal plane}\label{sec:PolRotationHorizontal}
The electron polarization vector precesses on the way through the accelerator to the experimental hall in the fields of the accelerator's bending magnets. The precession in an electromagnetic field due to Larmor and Thomas precession is described by the Thomas-BMT equation \cite{Bargmann:1959gz}. For the special case of vertical magnetic fields in the recirculating dipoles, only the horizontal components precess with a frequency $\omega = (1+a_e\gamma) \omega_{\mathrm c}$, where $\gamma$ is the Lorentz factor and $\omega_{\mathrm c}$ the cyclotron frequency \cite{Tioukine:2006ja}. Due to the anomalous magnetic moment $a_e$ of the electron, the polarization precesses faster than the momentum. Therefore the horizontal polarization orientation at the experiment substantially depends on the energy of the extracted beam.

The polarization vector can be rotated in the horizontal plane by using a Wien filter \cite{Tioukine:2006ja}, which is situated in the injection beam line. For conventional polarized experiments, the Wien filter is used to rotate the polarization vector so that it is aligned longitudinally in the experimental hall to maximize the sensitivity of experimental observables. The rotation angles of the polarization vector through the accelerator have been simulated for standard MAMI energies in \cite{SteigerwaldPhD}. In addition, the rotation angles have been determined for relevant beam energies by measuring the longitudinal beam polarization in the spectrometer hall as a function of different rotation angles of the Wien filter. Fitting these results with a cosine function yields the rotation angle of the polarization vector. The achieved accuracy is on the order of $2^\circ$. The rotation angles for the beam energies $\unit[570]{MeV}$ (the beam energy used during the $A_{\mathrm n}$ experiment) and $\unit[600]{MeV}$ (additional M\o ller measurements were performed at this energy) were determined to be $55^\circ$ and $189^\circ$, respectively.
\section{Rotation of the polarization vector into the vertical direction}\label{sec:PolRotationVertical}
\begin{figure}
  \centering
  \includegraphics[width=\linewidth]{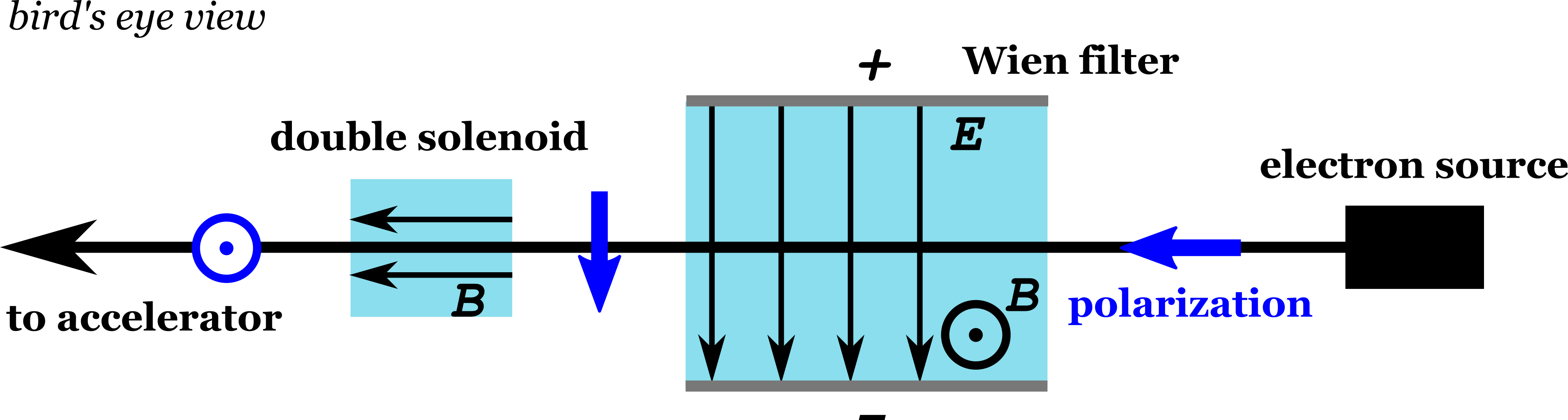} 
  \caption{Illustration of the steps to align the electron beam polarization vector vertically. The beam is initially emitted with longitudinal polarization. A Wien filter is used to rotate the polarization vector by $90^\circ$ in the horizontal plane. Subsequently a solenoidal field rotates the vector into the vertical direction.}
  \label{fig:PolRotationSource}
\end{figure}

The beam polarization vector can be rotated around the beam line axis by applying a solenoidal magnetic field aligned with the beam line axis. Hence, in the ideal case, one can first use the Wien filter to rotate the polarization vector to a transverse horizontal orientation, then apply a solenoidal field of appropriate strength to rotate it to the vertical direction. We have used a double solenoid which is located shortly behind the Wien filter, see Fig. \ref{fig:PolRotationSource}. 
It consists of two $\unit[46]{mm}$ long solenoids ($\unit[25]{mm}$ coils) with an aperture of $\unit[30]{mm}$. We have operated the two solenoids with the same polarity and with the same current to produce one continuous solenoidal field. The rotation angle of the polarization vector is then proportional to the integral of the longitudinal magnetic field.

Usually the magnets are operated in such a way that the fields are of opposite orientation and equal in strength, and serve as focusing elements with a focal length of approximately $\unit[24]{cm}$ inversely proportional to the integral of the squared longitudinal magnetic field. The polarity swap as well as a variation of the solenoid currents changes the focal length. Either the restoration of the original focal length could have been achieved with asymmetric currents in the double solenoid \cite{Steffens:1993ih}, or the effect could have been compensated by tuning of neighboring double solenoids. In practice, none of these methods was necessary. Specifically, during scans of the current in a limited range to optimize the double-solenoid setpoint, the change in focal length was small enough that the phase space of the beam still matched the acceptance of the injector linear accelerator.
\subsection{Depolarization effects}\label{sec:depolarization}
Fringe fields and inhomogeneities of the Wien filter and the solenoidal field could, in principle, result in a depolarization of the beam when electrons experience different fields due to the finite phase space volume of the beam.

However, to guarantee the same beam quality for every beamtime the same optimization procedure is performed routinely and results in a well centered beam in every single focusing element. For the small emittance of the MAMI beam at these locations of less than $\unit[1]{\pi\, mm\, mrad}$ and due to the fact that most fringe fields are very similar at the element entrance and exit, the effect is nevertheless negligible and it is not measurable with the MAMI polarimeters.

Depolarization further along the beam line is also negligible: in Refs. \cite{Steffens:1993ih,SteigerwaldPhD,Nachtigall:1998yy}, depolarization effects for a horizontally polarized beam were evaluated and they were found to be very small. For a vertical beam polarization alignment, the effects are even smaller due to the small horizontal magnetic field components that the beam experiences.  The vertical component is therefore conserved.

In particular, since the depolarization effects are negligible, the vertical polarization component can be deduced from a determination of the total beam polarization ``minus'' the measured residual horizontal polarization components.
\subsection{Consequences of misadjustments of the parameters}\label{sec:Suboptimal}
When the parameters of the Wien filter and the double solenoid are optimally adjusted, the polarization vector points in the vertical direction directly after the solenoids. Since there are no significant horizontal magnetic field components and depolarization effects, the beam polarization vector will remain vertically aligned throughout the transfer to the experimental hall with the same polarization degree.

A misadjustment of the parameters however results in significant residual polarization components in the horizontal plane behind the solenoids, thus a reduced vertical beam polarization component. Moreover, the horizontal components can give rise to false asymmetry contributions to experimental observables.

If the rotation angle of the Wien filter is slightly smaller than $90^\circ$, a small longitudinal polarization component remains. In the solenoidal field, that component is not changed. Due to subsequent polarization vector precession in the accelerator, that component will be rotated horizontally by $+55^\circ$ relative to the momentum direction, for the beam energy of the $A_{\mathrm n}$ experiment of $\unit[570]{MeV}$ (compare Sec. \ref{sec:PolRotationHorizontal}). For a Wien filter angle slightly larger than $90^\circ$, the corresponding polarization component points in the opposite direction in the experimental hall.

In case the field strength of the solenoids is slightly too small or too high to rotate the transverse horizontal component into the vertical direction, some polarization component remains in the transverse horizontal direction behind the solenoids and ends up in the spectrometer hall, with an additional relative rotation angle of $+55^\circ$ again.

Optimization of the Wien filter and double solenoid parameters is performed by considering measurements of the horizontal polarization components as described in detail in the next section.
\section{Vertical alignment procedure}\label{sec:PolMeasConcept}
The vertical polarization alignment can accurately be accomplished by minimization of the horizontal polarization components. These can be measured at two different locations: close to the source with a Compton \cite{Nillius:2011zz,Nillius:2016pns} and a Mott polarimeter and in the experimental hall by using the M\o ller polarimeter.

\begin{figure}
  \centering \includegraphics[width=0.95\linewidth]{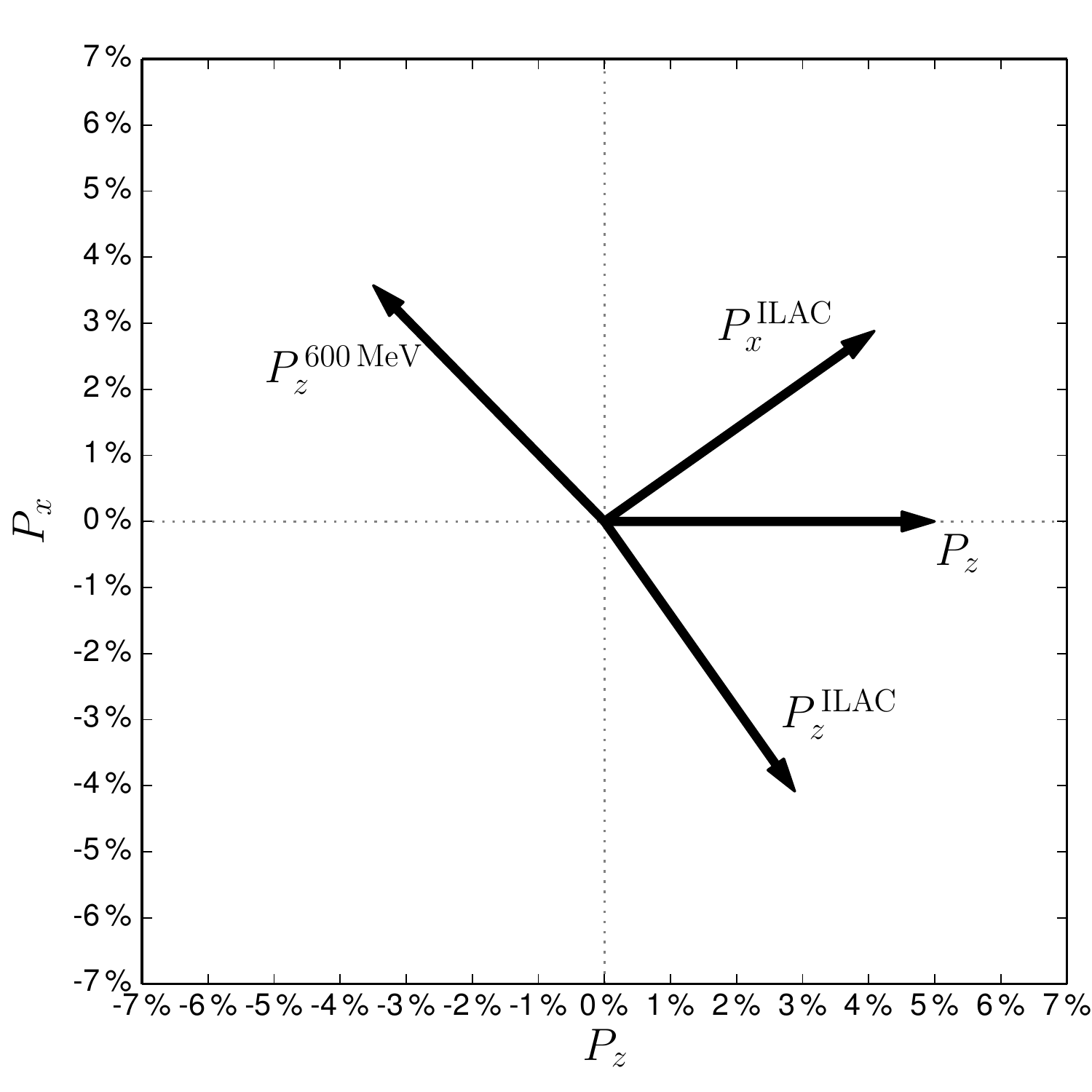}
  \caption{Overview of the polarization components that can be determined with the individual polarimeters. The M\o ller polarimeter is sensitive to $P_z$. With the Compton (Mott) polarimeter, the longitudinal (transverse horizontal) component can be measured behind the ILAC. Due to the precession in the accelerator, this component ends up rotated by $+55^\circ$ in the spectrometer hall, the corresponding axis is depicted as $P_z^{\mathrm{ILAC}}$ ($P_x^{\mathrm{ILAC}}$). The component along the axis labeled as $P_z^{\mathrm{\, 600\,MeV}}$ is measured with the M\o ller polarimeter, when the beam energy is temporarily set to $\unit[600]{MeV}$.
}\label{fig:A1_2d_Refs.pdf}
\end{figure}

The Compton and Mott polarimeters are positioned behind the injector linear accelerator (ILAC), where the $\unit[3.5]{MeV}$ electrons have already passed the Wien filter and the double solenoid. They are sensitive to the longitudinal and the transverse horizontal components, respectively. The Compton polarimeter is therefore well-suited to optimize the rotation angle of the Wien filter. With the Mott polarimeter, the applied field strength of the solenoids can be optimized. Due to the polarization precession in the accelerator, these components are related to the components in the experimental hall by a rotation, see Fig. \ref{fig:A1_2d_Refs.pdf}.

Let us denote the transverse horizontal, the vertical, and the longitudinal polarization components in the experimental hall by $P_x$, $P_y$, and $P_z$, respectively. $P_z$ can be measured directly with the M\o ller polarimeter. A possibility for inferring $P_x$ at the experiment is to temporarily change the final beam energy so that the rotation angle of the polarization vector through the accelerator is changed accordingly (compare Sec. \ref{sec:PolRotationHorizontal}). We used a beam energy of $\unit[600]{MeV}$, which is one of the standard energies of MAMI that can be selected by repositioning the beam extraction magnet \cite{Dehn:2016mqd}. In this case the relative rotation angle is $\approx 189^\circ$ (see Sec. \ref{sec:PolRotationHorizontal}), so the difference to the angle at $\unit[570]{MeV}$ is approximately $134^\circ$. This polarization component is labeled as $P_z^{{\mathrm{{\,600\,MeV}}}}$ in Fig. \ref{fig:A1_2d_Refs.pdf}.

For the purpose of the $A_{\mathrm n}$ experiment we performed the optimization with Mott and M\o ller polarimeter measurements at the beginning of the experiment and we used the Compton polarimeter for consistency check during the beamtime (Sec. \ref{sec:OfflineResidualPolarization}).

\begin{figure}
  \centering
  \includegraphics[width=\linewidth]{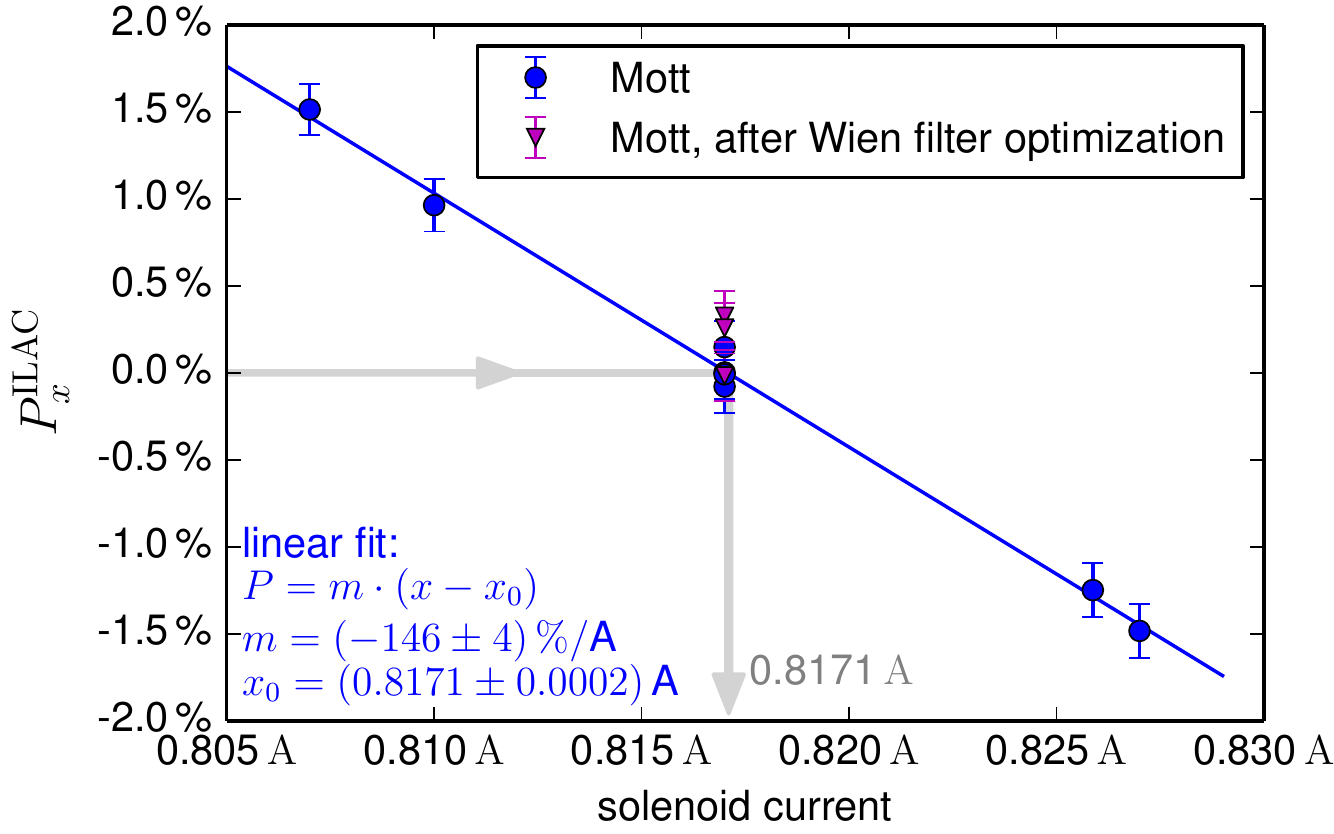}
  \caption{Measurement of the transverse horizontal polarization component with the Mott polarimeter in order to optimize the applied solenoid current (circles). Hereafter, a current of $\unit[0.817]{A}$ for the solenoids was used during optimization of the rotation angle of the Wien filter. With the optimized Wien filter angle, the triangular data points were measured. Only statistical errors are shown.} \label{fig:Mott_LineFit}
\end{figure}

We first performed the polarization measurements with the Mott polarimeter for different settings of the solenoidal field, with the Wien filter set to the nominal rotation angle of $90^\circ$, see Fig. \ref{fig:Mott_LineFit}.
We chose a current of $\unit[0.817]{A}$, for which the magnitude of the transverse horizontal polarization component behind the ILAC was minimal.

\begin{figure}
  \centering
  \includegraphics[width=\linewidth]{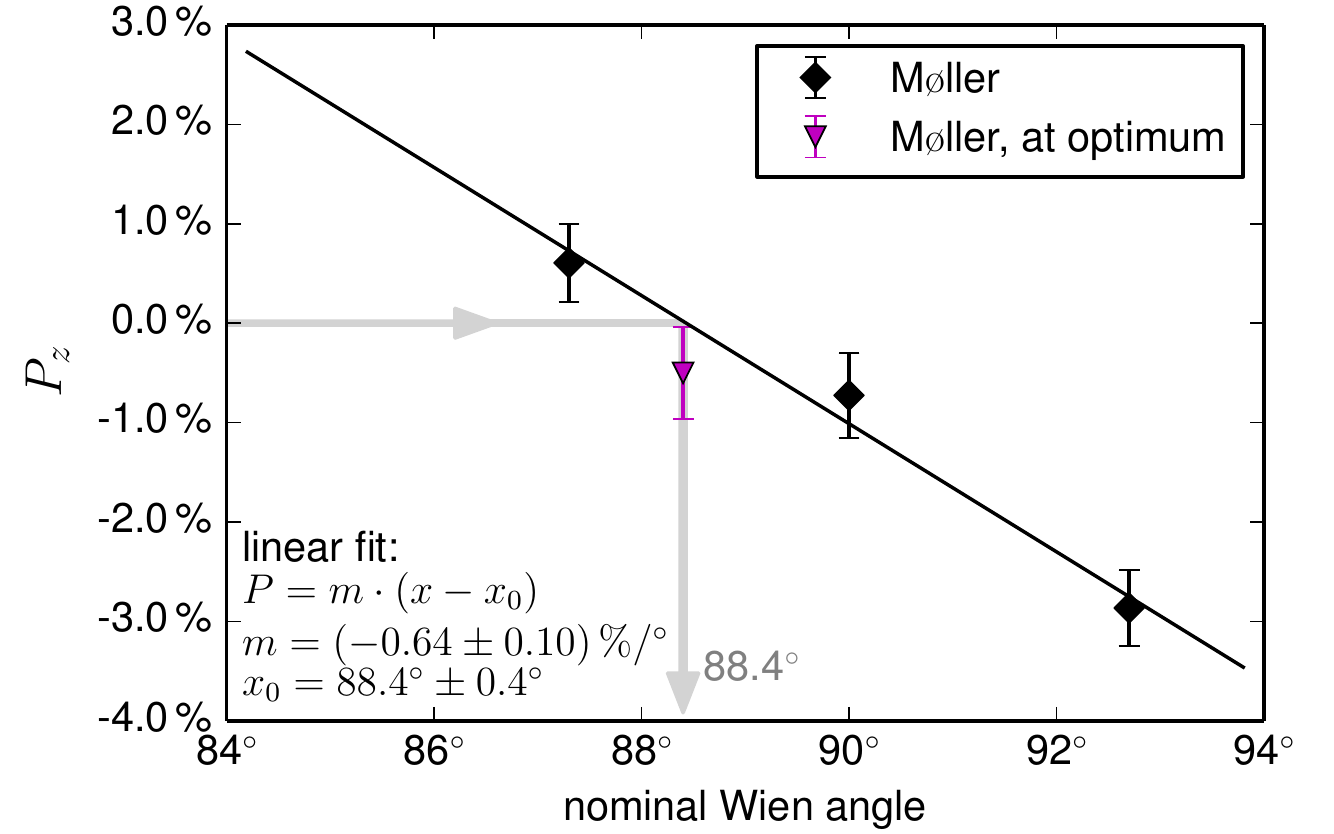}
  \caption{Measurement of the longitudinal beam polarization component in the spectrometer hall with the M\o ller polarimeter ($\unit[570]{MeV}$ beam energy) as a function of the nominal rotation angle of the Wien filter. 
According to the result of a linear fit on the diamond-shaped data points, a nominal Wien angle of  $88.4^\circ$  was chosen for the vertical polarization measurements. The triangular data point at the fit optimum was gathered directly afterwards. Only statistical errors are shown.}\label{fig:Moeller_LineFit}
\end{figure}

To minimize the second horizontal polarization component, we performed M\o ller measurements for different nominal Wien filter angles. The M\o ller polarimeter is sensitive to a linear combination of the horizontal components behind the linear accelerator. Given that one of these components has been already minimized by using Mott polarimeter measurements, the second component can be subsequently examined by using the M\o ller polarimeter. The results are shown in Fig. \ref{fig:Moeller_LineFit}.
We found $88.4^\circ$ as the optimum value for the nominal rotation angle of the Wien filter. 
With this new Wien filter setting, another measurement was performed with the Mott polarimeter (triangular data points in Fig. \ref{fig:Mott_LineFit}).
Due to the good agreement with the previous data points, these settings were used as final values for the beam-normal single-spin asymmetry measurements throughout the beamtime. 
\section{Systematic errors in the polarimeter measurements}\label{sec:Systematics}
The sensitivity of horizontal component measurements to a misalignment of the polarization vector is high, at the same time systematic effects can have a significant influence and deserve special attention.

Uncertainties related to factors entering the polarization determination like analyzing powers of the reactions, absolute target/absorber polarizations, unpolarized background, or necessary extrapolations to vanishing target thicknesses usually dominate the systematic errors. However, these are negligible when small polarization components are measured, because the relative uncertainties of these sources do not change, but the absolute uncertainties become small compared to the statistical precision.

In contrast, false asymmetries do not scale with the size of the measured polarization components and become dominant when the ``true'' polarization components are small. Examples are an inaccurate helicity-dependent luminosity determination used for normalization of the raw polarimeter count rates, a slight misalignment of the polarimeter setup, contributions from another analyzing power component of the scattering process utilized for the polarimetry, or a helicity-dependent beam position on the polarimeter target resulting in an additional false asymmetry of the count rates.

To get a handle on the systematics, we took data dedicated to a direct estimate of false asymmetries. Additionally, we compared the data from the different polarimeters, which have different systematics.

To estimate false asymmetries one needs to perform precise asymmetry measurements of known asymmetries. One might use for instance an unpolarized beam, an unpolarized absorber (Compton polarimeter), or an unpolarized target (M\o ller polarimeter). 
\subsection{False asymmetry estimate of M\o ller polarimeter measurements}\label{sec:FAMoeller}
\begin{figure*}
  \centering \includegraphics[width=\linewidth]{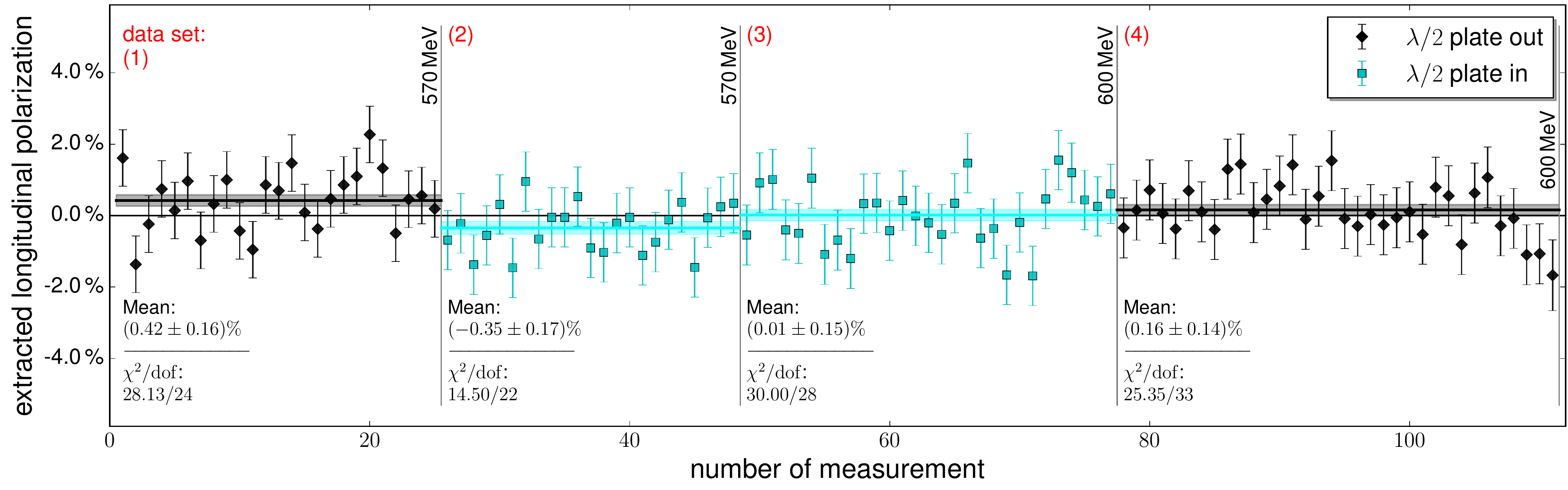} 
  \caption{Results of M\o ller measurements on a copper foil. The results are used to estimate the contribution of false asymmetries. Data sets 1 and 2 have been taken at a beam energy of $570\,$MeV, 3 and 4 at $600\,$MeV.}\label{fig:CopperZero.pdf}
\end{figure*}

Performing measurements with the M\o ller polarimeter with an unpolarized target can be used to estimate false asymmetries, since the expected asymmetry is zero. Depolarizing the standard iron foil might be hard to achieve with the desired accuracy and raises the problem that the target magnetic field is finally turned off, thus slightly changing the trajectories of the electrons. Therefore, we used a copper foil instead. The beam polarization was oriented vertically during these measurements. Since copper is diamagnetic, vanishing asymmetries are expected. The size of false asymmetries can then be estimated from the deviation of the measurement from zero. Measurements were performed for both relevant beam energies ($570\,$MeV and $600\,$MeV), and for both $\lambda/2$ plate states. The results are shown in Fig. \ref{fig:CopperZero.pdf}. For better comparison, we converted the observed asymmetries to apparent polarization values that would be extracted if these asymmetries were measured with the commonly used ferromagnetic target (i.e., we multiplied the observed asymmetries by a factor of $\approx 16$, which corresponds to a division by the experimental analyzing power and the effective target electron polarization of the iron target).
The averaged result of the data for $\unit[570]{MeV}$ with the $\lambda/2$ plate out significantly differs from a zero measurement, which is a clear indication for false asymmetries. Furthermore, it is incompatible with the result of the data for the same beam energy, but with the $\lambda/2$ plate inserted: A two-sample $t$-test for the null hypothesis that these two data samples originate from the populations with equal means (assuming that they have identical variances) gives a $p$-value of $0.0013$. This strongly indicates that the two population means are not equal. It is more likely that the means are equal in magnitude but have a different sign ($p=0.73$). This is an indication that the dominant part of the false asymmetries is related to beam properties (like helicity-dependent beam positions or aberrations) rather than to polarimeter issues. However, the deviation from zero is quite small (e.g., $0.42\,\%$ absolute for the $\lambda/2$ plate out data set, which corresponds to an instrumental asymmetry of $0.026\,\%$), and the result for $600\,$MeV is even compatible with zero. Based on these results, we use the deviations from the zero-asymmetry measurements and their precisions (individually for both $\lambda/2$ plate states and beam energies) for a false asymmetry correction and its corresponding systematic error.
\subsection{False asymmetry estimate of Mott polarimeter measurements}\label{sec:FAMott}
To estimate false asymmetry contributions to the Mott polarimeter measurements, auxiliary measurements have been performed with longitudinally polarized electrons at the Mott target. Since the Mott polarimeter is not sensitive to the longitudinal polarization component, such measurements should result in a vanishing asymmetry.

Longitudinally polarized electrons are guided directly to the Mott polarimeter when the Wien filter and the solenoids are not used to rotate the polarization. In a strict sense, the polarization orientation is, even in this case, not exactly longitudinal at the Mott target. An $\alpha$-magnet at the very beginning of the electron beam line bends the $\unit[100]{keV}$ beam by $270^\circ$ and results in a polarization vector rotation of approximately $0.37^\circ$. Therefore there is a small polarization component in the vertical direction, to which the Mott polarimeter is not sensitive. However, two dipole magnets in the ILAC beam line that are used to bend the beam by $30^\circ$ to the Mott polarimeter setup, also cause a slight rotation of the beam polarization in the horizontal plane by $0.27^\circ$ relative to the beam direction. The expected polarization component measurable by the Mott polarimeter is then approximately $0.005 \cdot P_{\mathrm{total}}$.

We measured a residual polarization of $-0.41\%\pm 0.07\%$ ($0.39\% \pm 0.07\%$) for the $\lambda/2$ plate not inserted (inserted), whereas the expectation due to the dipole bending is $0.39\%$ ($-0.39\%$). We have repeated such measurements in a following beam test, the results were similar. To investigate the source of this deviation, we performed additional systematic tests. After all, the origin of the inconsistency remains unclear. The present estimate for the systematic uncertainty is $0.8\%$ absolute (this value matches the reported deviation). Clarifying this issue thus reducing the magnitude of the total systematic error associated with the polarization measurement is beyond the scope of this paper, but thorough measurements will be performed in the near future to understand the deviation.
\section{Determination of the polarization orientation}\label{sec:OfflineResidualPolarization}
In order to reconstruct $P_x$ and $P_z$ and to draw a quantitative conclusion about the polarization orientation, different types of polarization measurements are performed and combined.
\subsection{Measurements with optimized settings}\label{subsec:OfflineResidualPolarization_OptimizedOut}
\begin{figure*}
  \centering 
  \includegraphics[width=0.495\linewidth]{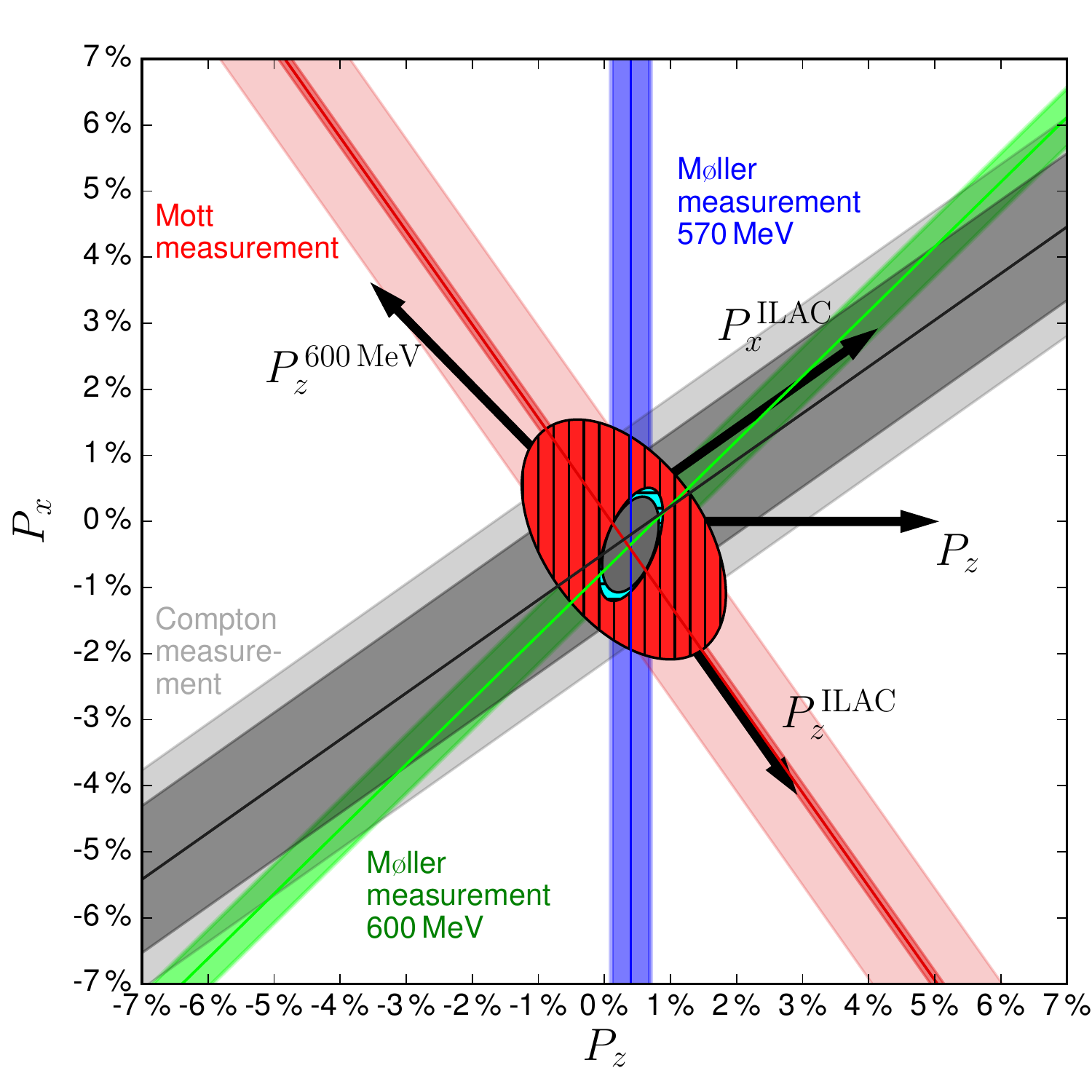}
  \includegraphics[width=0.495\linewidth]{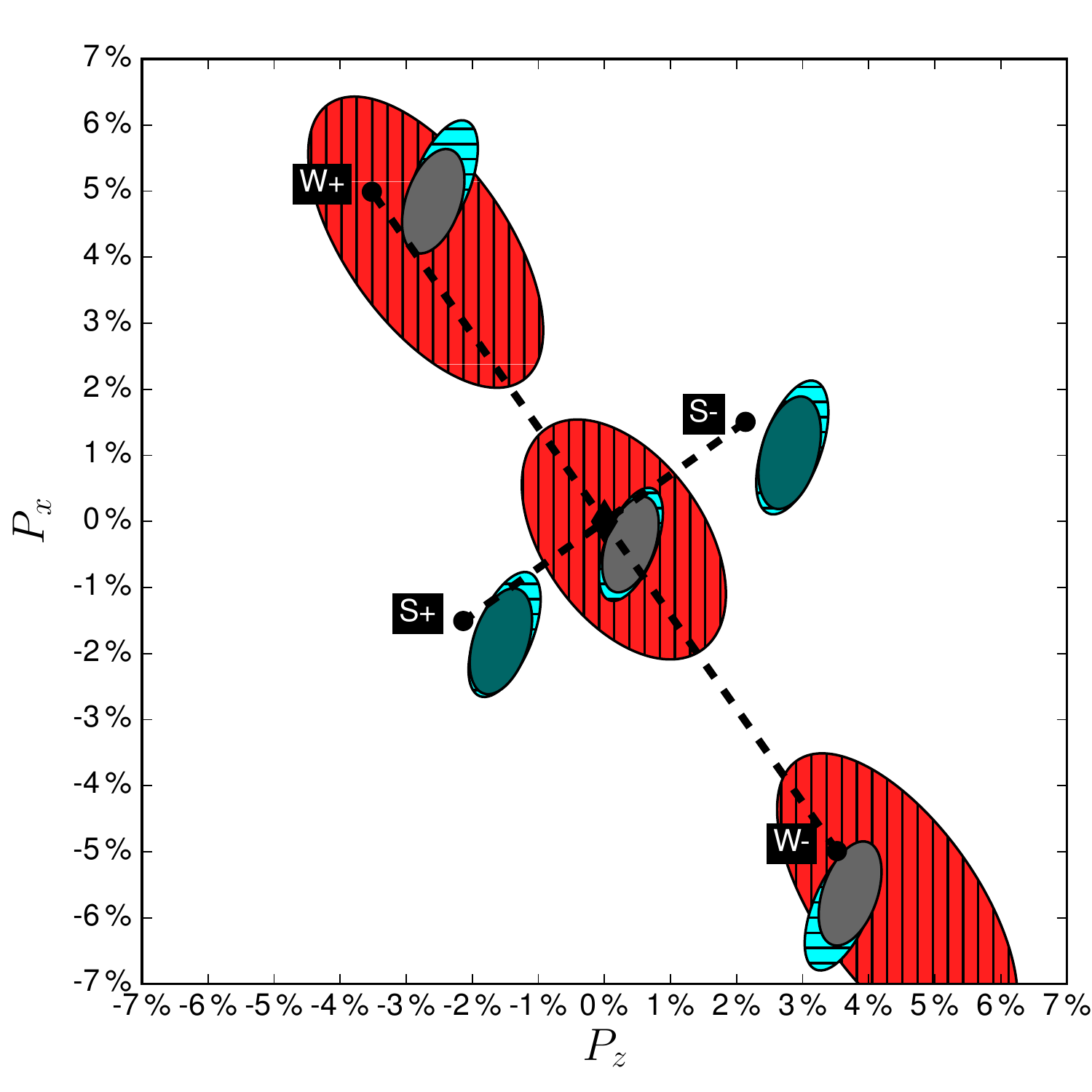}
  \caption{{\bf Left:} Reconstructed horizontal beam polarization components for the optimized Wien filter and solenoid settings. The main component of the polarization points in the $y$-direction for these settings. The result of a maximum likelihood fit when combining all available polarimeter data is shown as solid ellipse. Consideration of only the Mott and Compton result gives the vertically hatched, combination of the two M\o ller measurements the horizontally hatched ellipse. {\bf Right:} Fit results for polarimeter data with slightly changed Wien filter and solenoid settings: The circles with labels $\mathrm S+$ ($\mathrm S-$) correspond to the expectation when the solenoid current is increased (lowered) by 2\,\%, as it was done for these measurements. The circles with labels $\mathrm W+$ ($\mathrm W-$) show the expectations when the rotation angle of the Wien filter is enlarged (reduced) by 5\,\%.}\label{fig:A1_2d_CentralWithLines}
\end{figure*}

The results for the optimized Wien filter and double solenoid configurations are depicted in Fig. \ref{fig:A1_2d_CentralWithLines} (left). The M\o ller polarimeter measurements performed at the beam energy of the $A_{\mathrm n}$ experiment ($\unit[570]{MeV}$) are sensitive to $P_z$. The measurement result is represented as a line perpendicular to the $P_z$ axis with corresponding error bands. Here, the correction mentioned in Sec. \ref{sec:FAMoeller} has been applied. The results of the Compton and Mott polarimeter measurements as well as the one deduced from the M\o ller measurement at a beam energy of $\unit[600]{MeV}$ are displayed similarly. The inner and outer bands correspond to the statistical and the statistical and systematic errors added in quadrature, respectively. For the M\o ller measurements, the two bands are hardly distinguishable since the statistical uncertainty is considerably larger than the estimated systematic error.

To estimate $P_z$ and $P_x$ simultaneously from the different polarimeter measurements, we used a maximum-likelihood fit with inputs the measured results and the polarization vector precession angles through the accelerator. Errors considered in the fitting procedure are the quadratically added statistical and systematic uncertainties of the individual measurements. An additional uncertainty enters through the imperfect knowledge of the polarization vector rotation angles through the accelerator.
The influence has been checked by varying these angles in the limits of their uncertainties and was found to be negligible.

The result of a fit considering only the Compton and the Mott polarimeter measurements is represented by the vertically hatched ellipse (the $1\sigma$ error ellipse corresponding to the estimated covariance matrix). Fitting the two M\o ller polarimeter results yields the horizontally hatched (blue) ellipse. Despite the different errors, the fit results are very similar. Finally, the result of a fit considering all available polarimeter measurements is shown as the solid ellipse in the figure. 
The results for $P_z, P_x,$ and the covariance matrix $C$ are
\begin{eqnarray*}
(P_z, P_x) &=& (0.39\,\%, -0.35\,\%),\\
C &=& 
\begin{bmatrix}
\phantom{+}(0.28\,\%)^2 & +(0.24\,\%)^2 \\
+(0.24\,\%)^2 & \phantom{+}(0.48\,\%)^2 
\end{bmatrix}. \\
\end{eqnarray*}
With a total polarization of approximately $79\,\%$ at the end of the beamtime, this corresponds to an angle misalignment of merely $0.38^\circ$. Therefore, the degree of vertical polarization differs only marginally from the total degree of polarization ($(P_{\mathrm{total}}-P_y)/P_{\mathrm{total}}=2\cdot 10^{-5}$).
\subsection{Variation of the Wien filter and solenoid settings}
We also tested our underlying assumption about how the Wien filter and the solenoidal fields rotate the polarization orientation. In Fig. \ref{fig:A1_2d_CentralWithLines} (right), the same ellipses as in the left panel are shown.  In addition, the fit results for measurements with changed settings of the Wien filter and the solenoids are drawn. Either the nominal Wien filter angle (optimized value: $88.4^\circ$) was varied by $\pm 4.3^\circ$, or the current of the solenoids (optimized value: $0.817\,$A) was changed by $\pm 0.017\,$A. Compton polarimeter measurements were not performed for the latter settings. The individual results approximately coincide with the expectations (circles labeled by $\mathrm S+,\mathrm S-, \mathrm W+, \mathrm W-$).
\section{Stability during the beamtime}\label{sec:HorizontalResultAndStability}
\begin{figure}
  \centering \includegraphics[width=\linewidth]{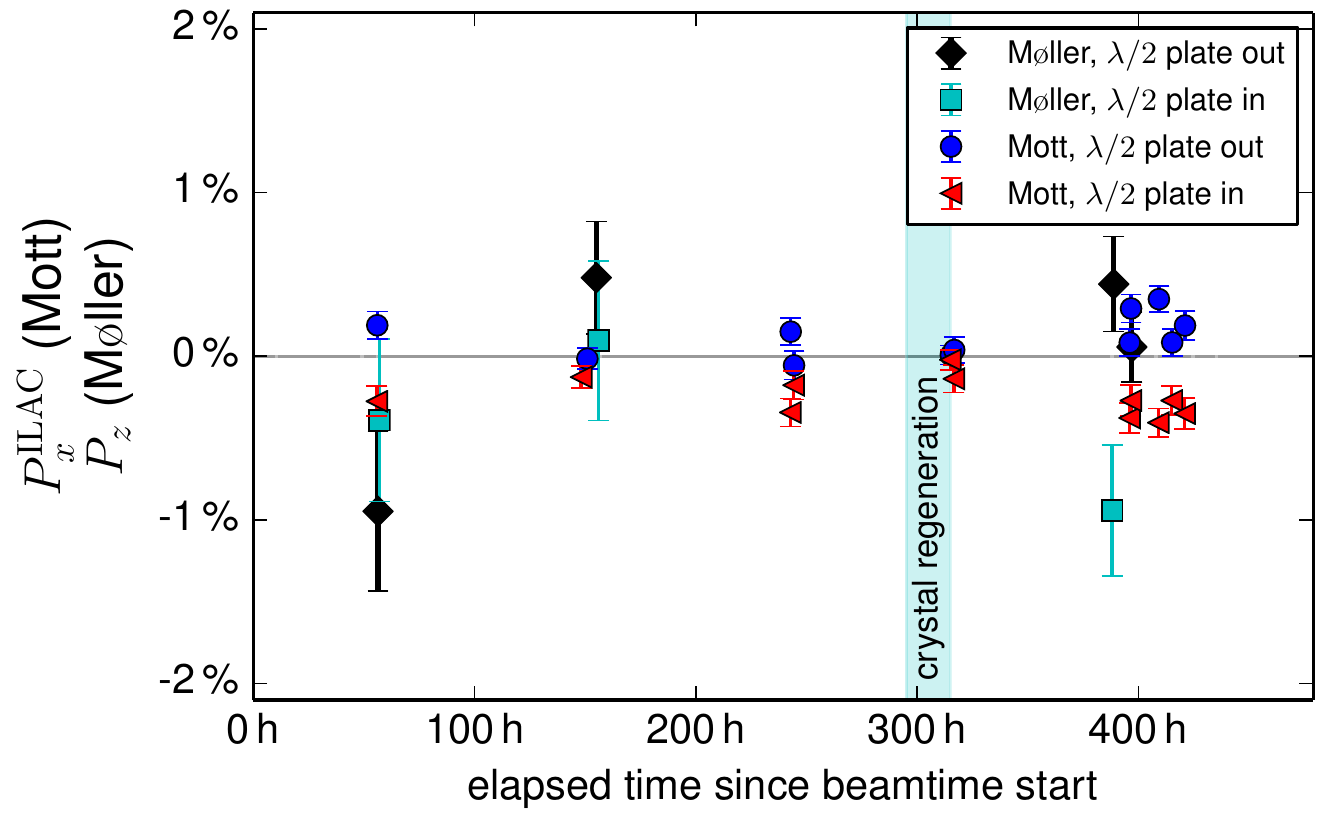}
  \caption{Measurements of residual polarization components during the beamtime with the Mott polarimeter and the M\o ller polarimeter at 570\,MeV beam energy. Only statistical errors are shown.}\label{fig:ResidualStability}
\end{figure}

To check the stability of the polarization alignment on the three-week time scale of the experiment, several measurements with the Mott polarimeter and the M\o ller polarimeter at $\unit[570]{MeV}$ beam energy have been performed throughout the beamtime, see Fig. \ref{fig:ResidualStability}.
The agreement between the individual Mott measurements is very good. There is a systematic difference between data collected with and without the $\lambda/2$ plate inserted. A small detune of the double solenoid current would cause exactly such an effect. The magnitude of the systematic errors associated with these measurements, as outlined in Sec. \ref{sec:FAMott}, hampers a verification of such an assumption.

There is a larger spread of data values for the M\o ller polarimeter measurements. However, these data do not indicate variations of the residual horizontal polarization components above the $1\,\%$ level. The corresponding uncertainties for the analysis of the $A_\mathrm{n}$ experiment are sufficiently small compared to the achieved statistical precision \cite{An2016}.
\section{Summary}\label{sec:Summary}
The $A_{\mathrm n}$ experiment of the A1 collaboration was dedicated to the measurement of the beam-normal single-spin asymmetry in elastic scattering from $^{12}\mathrm C$. For this asymmetry measurement, the beam polarization was rotated from an initial longitudinal to a vertical orientation. A Wien filter spin rotator (located in the injector linear accelerator beam line) has been used to first rotate the polarization into the transverse horizontal plane. Afterwards the polarization was rotated into the vertical direction by means of a solenoidal field of appropriate strength.

In order to measure the total beam polarization, to achieve a good vertical beam polarization alignment and to provide an estimate of remaining horizontal beam polarization components, numerous polarimeter data have been collected. The absolute beam polarization (around $80\,\%$ throughout the beamtime) was determined with a M\o ller polarimeter by using the most favourable beam energy for that measurement. The variation of the polarization degree with time was measured with a Mott polarimeter.

To optimize the solenoid current and the Wien filter angle, polarization measurements with the Mott and the M\o ller polarimeter served as benchmark by providing results for the horizontal polarization components to be minimized. Dedicated tests have been performed to confirm the expectation for the effect of the Wien filter and solenoid settings on the polarization orientation. To allow a quantitative estimate of residual horizontal polarization components, not only the results from the Mott polarimeter and the M\o ller polarimeter at the beam energy of the $A_{\mathrm n}$ experiment have been performed, but these were supplemented by M\o ller polarimeter measurements at a slightly different beam energy and by Compton polarimeter measurements. Auxiliary polarimeter measurements have been performed to estimate the contribution of false asymmetries, which are the dominant error sources when measuring small polarization components. Combining these measurements yields horizontal polarization components, which are indeed close to zero, with uncertainties on the $1\,\%$-level. Related uncertainties in the extraction of the beam-normal single-spin asymmetry are reasonably small. The newly developed method for determining the beam polarization components by using combined measurements of three different polarimeters will allow future precision studies with polarized beam at MAMI.
\section*{Acknowledgement}
This work was supported in part by the Deutsche Forschungsgemeinschaft with the Collaborative Research Center 1044, the PRISMA (Precision Physics, Fundamental Interactions and Structure of Matter) Cluster of Excellence, BMBF Verbundforschung within project 05H12UM6 "Spin Optimierung", and the Federal State of Rhineland-Palatinate.
\section*{References}
\bibliographystyle{elsarticle-num} 
\bibliography{cite}
\end{document}